\documentclass[12pt]{article}
\usepackage[utf8]{inputenc}
\usepackage[margin=1in]{geometry}
\usepackage[square,numbers,comma,sort&compress]{natbib}
\usepackage{textcomp,graphicx,bm,amsmath,xcolor}
\usepackage{amssymb,color,caption,subcaption,setspace}
\usepackage[square,numbers,comma,sort&compress]{natbib}

\usepackage{setspace}   
\usepackage{lipsum} 
\usepackage{wrapfig}

\usepackage{authblk}
\title{Dissipation engineering in metamaterials by localized structural dynamics}
\author{Cl\'emence L. Bacquet}
\author{Mahmoud I. Hussein}
\affil{Ann and H.J. Smead Department of Aerospace Engineering Sciences, University of Colorado Boulder, Boulder, Colorado 80309, USA}
\affil{Corresponding author: M. I. Hussein (mih@colorado.edu)}

\date{\today}                     

\begin{document}
	
\maketitle


\begin{abstract}
In civil, mechanical, and aerospace engineering, structural dynamics is commonly understood to be a discipline concerned with the analysis and characterization of the vibratory response of structures. Key elements of the response are the amplitude, phase, and damping ratio, which are quantities that vary with the excitation frequency. In this paper, we extend the discipline of structural dynamics to the realm of materials engineering by intrinsically building localized substructures within, or attached to, the material domain itself$-$which is viewed as an extended medium without defined external boundaries. Our system is essentially a locally resonant elastic metamaterial, except here it is viewed from the perspective of unique dissipation characteristics rather than subwavelength effective properties or band gaps, as widely done in the literature. We provide a theory, validated by experiments, for substructurally synthesizing the dissipation under the conditions of free-wave motion, i.e., waves not constrained to a prescribed driving frequency. We use an extended elastic beam with attached pillars as an example of a metamaterial. When compared to an identical infinite beam with no attached substructures, we show that within certain frequency ranges the metamaterial exhibits either enhanced or reduced dissipation$-$which we refer to as positive and negative metadamping, respectively. These regimes are rigorously identified and characterized using the metamaterial's band structure and wavenumber-dependent dissipation diagram. This theory impacts applications that require a combination of high stiffness and high damping or, conversely, applications that benefit from a reduction in loss without the need to change the backbone constituent material.
\end{abstract}




\section*{Introduction}
The dynamic response of a structure depends primarily on the properties of the materials from which it is formed. Important elements of the response are the amplitude, phase, and damping ratio (loss factor), which are all quantities that vary with the excitation frequency.~The aim of vibration engineering is to select the most appropriate materials and use these materials to design a structure with the desired response characteristics~\cite{Inman_2013,Rao_2016}.~In numerous applications, such as support ribs in aircraft wings, the goal is to arrive at a material-structure solution that not only minimizes the amplitude of vibration (at a frequency or a range of frequencies of operation) but also retains sufficient stiffness to avoid excessive elastic deformation that would undermine the function of the structure and also possibly reduce its lifespan. In this class of problems, it is desirable to use materials that exhibit simultaneously high stiffness and high damping capacity~\cite{Chen_Lakes_1993}.~This has been pursued using composite materials, as surveyed in a recent review by Treviso et al.~\cite{Treviso_2015}.~Strategies to achieve this target include embedding microscopic inclusions exhibiting negative stiffness~\cite{lakes_extreme_2001}, introducing a hierarchical layout of constituents~\cite{Lakes_2002}, and optimizing the distribution of constituents~\cite{Andreassen_2014}. In another class of problems, the goal is to reduce loss in order to enhance signal generation or transmission, e.g., in terms of resonance quality factor and distance traveled, respectively. Acoustic devices based on microelectromechanical systems (MEMS) is an example of an application where minimum dissipation is desired.~In both categories, whether the desire is to increase or decrease dissipation at the structural or device level, one is limited to what is provided by the intrinsic damping properties of existing materials~\cite{Ashby_1989}. This leaves the design process with only two options: optimal materials selection~\cite{Farag_2013,Ashby_2015} and structural/device engineering~\cite{Inman_2013,Rao_2016}.~In this work, we introduce a third option:~\it dissipation engineering \rm within the material itself. \\

When presented with an engineering material in its raw form, i.e., without the proposed intervention of dissipation engineering, damping is generated primarily by thermoelastic effects which are influenced typically by the following underlying features~\cite{Lakes_2009}: (1) microstructural atomic configuration (e.g., crystal versus non-crystal; single crystal versus polycrystal), (2) presence of defects at the microstructural level (e.g., grain boundaries, dislocations, vacancies), and (3) rheological properties (e.g., fluid-like behavior as in rubbery polymers).~Needless to say, all these features may be altered by intervention at the microscopic level.~For example, heat treatment enables recrystallization and alloying pins dislocations$-$two outcomes that reduce loss~\cite{Ashby_1989}.~However, such treatment routines essentially redefine the source material itself and merely transitions it from one form to another from among the group of existing materials$-$each with its pertinent trade-offs in mechanical and other physical properties.~Our proposition, on the other hand, is to engineer the dissipation at the "macroscopic level"\footnote{Dissipation engineering at the macroscopic level is additive; it builds on what may be achieved at the micrscopic level.~Furthermore, it enables increased, or decreased, dissipation at very low frequencies, i.e., frequencies far lower than what can be accessed by microscopic mechanisms.} and in doing so create an elastic (or acoustic) metamaterial~\cite{liu_locally_2000, Hussein_AMR_2014} with either enhanced or reduced damping capacity within sets of predetermined frequency ranges. We refer to this concept as \it metadamping \rm \cite{hussein_metadamping_2013}.\\

Metadamping has been realized by introducing locally resonant substructures within a host material~\cite{hussein_metadamping_2013,frazier_viscoelastic_2015,Chen_2016, depauw_metadamping_2018}, and in some cases utilizing other physical mechanisms such as embedding an oscillator with negative stiffness~\cite{antoniadis_hyper-damping_2015}.~The focus has mainly been on positive metadamping where the level of dissipation is enhanced compared to an equivalent material that does not include oscillating substructures\footnote{The "equivalence" may be established by ensuring that the metamaterial is compared to another material with identical overall unit-cell mass and static stiffness~\cite{hussein_metadamping_2013}.}.~Furthermore, metadamping was previously presented only in theory and mostly in the context of overly simplified toy models based on masses, springs, and dashpots.~Here, we demonstrate the concept in a real-world experimental set up in the form of an extended all-aluminum beam with pillars periodically branching out from one of the beam's surfaces.~The pillars serve the role of the resonating substructures.~Our choice of a medium with one-dimensional periodicity is only for ease of exposition as the underlying dynamical behavior we are interested in takes place regardless of the dimensionality of the periodicity.~Also, and very importantly, we extend our investigation to additionally encompass the phenomenon of negative metadamping, i.e., where the metamaterial becomes less lossy than the reference configuration.~First, we use experiments and corresponding brute-force numerical simulations applied to a finite-sized pillared beam to provide evidence of metadamping.~We then present a new theoretical measure of metadamping that is applicable to experimentally realizable metamaterial configurations.~This measure is derived from a formal Bloch waves~\cite{bloch_uber_1929} analysis of the unit cell where the wavenumber-dependent damping ratio of the pillared beam is compared to that of a uniform beam with asymptotically converging damping ratios.\\

\section*{Results and Discussions}
\begin{figure}[t]
	\centering
	\includegraphics[width=87mm]{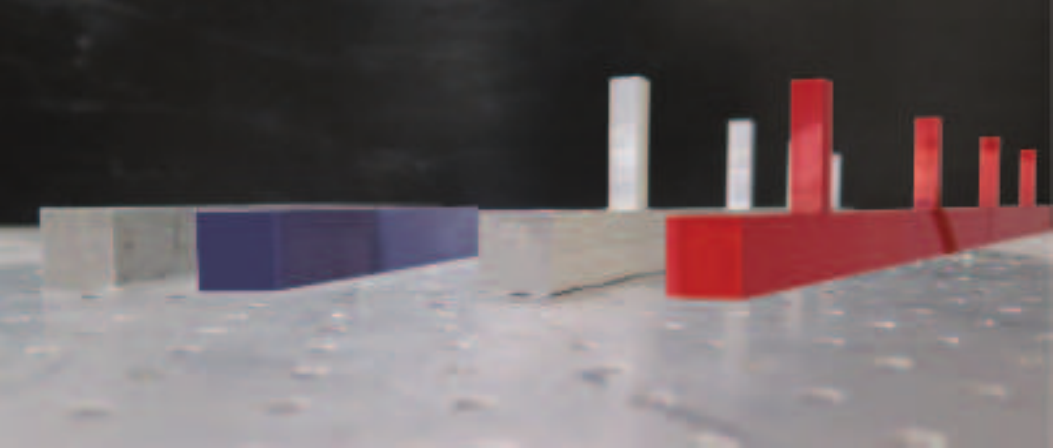}
	\caption{\small{\textbf{Test samples and models.} Photograph of the experimental uniform (left) and pillared (right) beams and their corresponding numerical models. The backbone beam represents an extended 1D elastic material and the pillars represent add-on mechanical substructures. Combining these two "parts" yields the "whole" which is an elastic metamaterial with emergent damping (metadamping) properties~\cite{hussein_metadamping_2013} at specific frequency ranges.~This system is a realization of the concept of building a material out of structures rather than a structure out of materials~\cite{Hussein_Blog_2012}.}}
	\label{fig:Fig1}
\end{figure}

\begin{figure}[b!]
	\centering
	\includegraphics{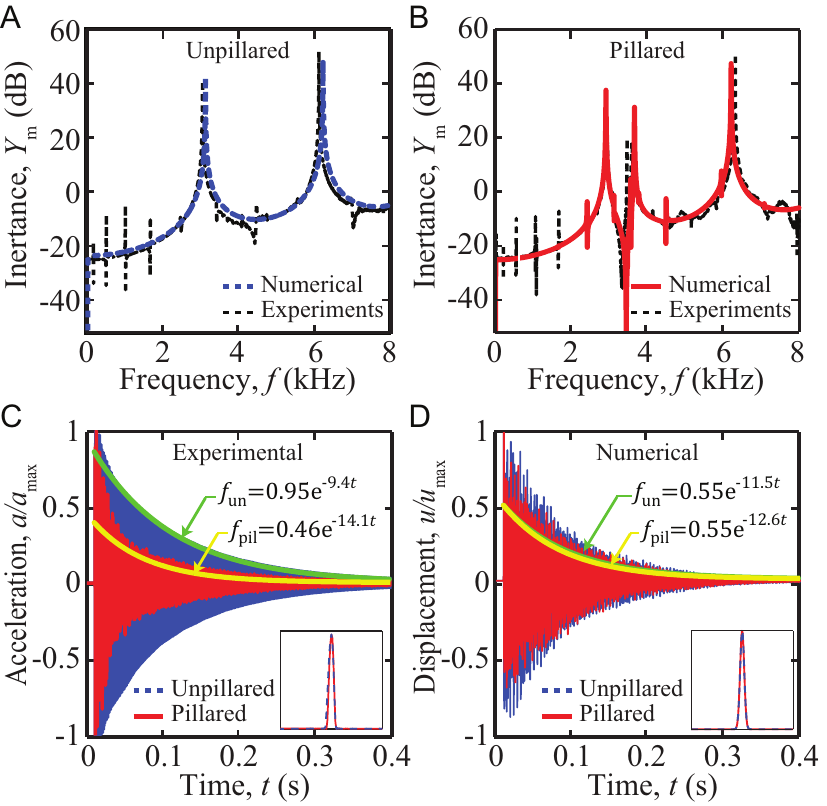}
	\caption{\small{\textbf{Finite-structure response: Evidence of metadamping.} Numerical and experimental frequency response functions for the (\it{A}\rm) unpillared and (\it{B}\rm) pillared beam. (\it{C}\rm) Experimental and (\it{D}\rm) numerical temporal responses for the two structures.~The yellow and green exponential curves are fitted to the unpillared and pillared time signals, respectively, to determine their time decay rates.~The spatial profiles of the impulse excitation are shown in the insets.}}
	\label{fig:Fig2}
\end{figure}

Figure~\ref{fig:Fig1} illustrates the two test structures we consider: an unpillared beam and a pillared beam, both entirely made out of aluminum.~The pillared beam consists of four periodic unit cells laid out along the axial direction, where each comprises a square-section base beam component and a square pillar standing on the top.~The pillar is milled into shape to ensure seamless connectivity.~The unpillared beam has the same cross section and is used as a reference case.~In Figs.~\ref{fig:Fig2}A and~\ref{fig:Fig2}B, experimental frequency response functions (FRF) for longitudinal vibrations are plotted for the unpillared and pillared beams, respectively (see Supplementary Information for details on the experimental procedure).~The temporal response to a particular profile of impulse excitation is shown in Fig.~\ref{fig:Fig2}C for the same test structures.~To identify the presence of metadamping and quantify its intensity, we compare the rate of time decay between the two beams.~The procedure is as follows. First, the first derivative of the displacement response is evaluated.~Then, the maxima, which are the points where the derivative changes sign, are extracted.~Exponential functions of the form~\(f(t)=ae^{-bt}\) are then curve-fitted to these extracted response peaks, where the exponential decay constant \textit{b} provides a direct measure of the degree of dissipation. \\

\indent \bf Evidence of metadamping \rm We define the ratio~\(r=b_{\rm pil}/b_{\rm un}\), where ``pil'' and ``un'' denote pillared and unpillared, respectively, as a metric for metadamping.~A ratio greater than unity signifies positive metadamping, i.e., the time response of the pillared beam decays faster than that of the unpillared beam.~Conversely, a ratio less than unity indicates negative metadamping. Here, we report an experimental metadamping ratio~\(r_{\rm exp}=1.49\) which is indicative of nearly 50\% positive metadamping, i.e., the pillared beam exhibits 50\% higher dissipation than the unpillared beam for this particular form of excitation.~To support these experimental findings, we perform a numerical experiment using an in-house, finite-element (FE) code~\cite{Hussein_PRSA_2009}.~We use a viscoelastic damping model of the form~\cite{wagner_symmetric_2003,hussein_damped_2013,frazier_viscoelastic_2015}
\begin{equation}
\mathbf{M \ddot{u}}(t)+ \int_{k=1}^{t} \mu e^{-\mu(t-\tau)} \mathbf{C \dot{u}}(\tau)d\tau+\mathbf{Ku}(t)=\mathbf{f}(t),  \label{EOM}
\end{equation}
where, for simplicity, the damping matrix~\textbf{C} is assumed to be proportional to the mass \textbf{M} and stiffness \textbf{K} matrices, such that~\(\mathbf{C}= p\mathbf{M}+q\mathbf{K}\) (see Supplementary Information for information on FE matrices).~The parameters \(p\) and \(q\), as well as a relaxation parameter \(\mu\), are determined by a unique curve-fitting procedure described below. In Figs.~\ref{fig:Fig2}A and \ref{fig:Fig2}B, the numerical FRFs are shown to agree very well with their experimental counterparts.~Once the damping parameters are selected, we implement a time-integration scheme particularly suited for exponentially damped systems~\cite{adhikari_direct_2004} and obtain the temporal response shown in Fig.~\ref{fig:Fig2}D (see Supplementary information for details on these numerical calculations).~We report a numerical metadamping ratio~\(r_{\rm num}=1.09\). While the degree of the predicted metadamping is lower than the experimental results, this confirms the pillared beam exhibits higher dissipation for the applied excitation profile.\\

\begin{figure}[t!]
	\begin{center}
		\includegraphics{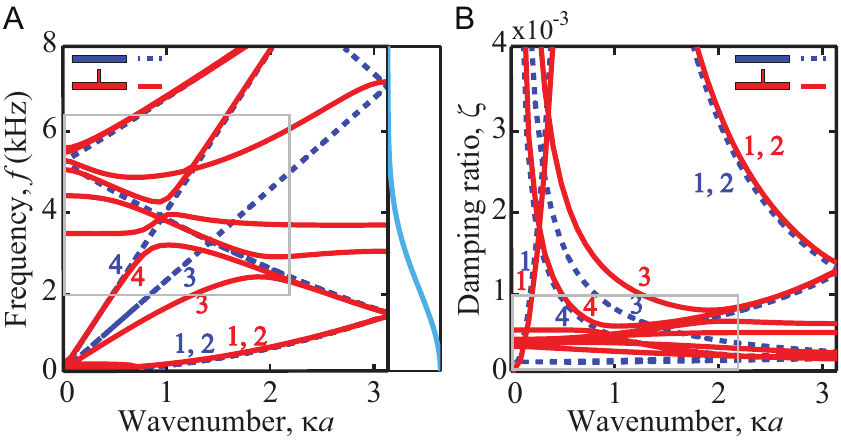}
		\caption{\small{\textbf{Unit-cell analysis: Elucidation of metadamping.} (\it{A}\rm) Dispersion diagram for the unpillared (dashed blue) and pillared (red) unit cells. The first four branches corresponding to the flexional, torsional, transverse, and longitudinal modes of the beam are identified~\cite{khales_evidence_2013}.~The frequency content of the impulse excitation is depicted in the inset.~(\it{B}\rm) Damping-ratio diagram with the first four damping branches identified.~The boxes in (\it{A}\rm) and (\it{B}\rm) outline the close-up views shown in Fig.~\ref{fig:Fig4}.}}
		\label{fig:Fig3}
	\end{center}
\end{figure}

\indent \bf Dispersion and damping ratio curves \rm In elastic metamaterials such as our pillared beams, local resonances appear as horizontal lines in the frequency versus wavenumber diagram and couple with the dispersion curves of the underlying medium.~These couplings may generate band gaps, which are frequency ranges characterized by strong spatial attenuation.~Figure~\ref{fig:Fig3}A displays the elastic band structure of both the pillared and unpillared beams obtained by applying Bloch's theorem~\cite{bloch_uber_1929} on the unit cell following~\eqref{EOM} and solving for the wavenumber-dependent complex frequencies~\(\lambda_s(\kappa)\).~The phenomenon of resonance hybridization is clearly evident in this diagram (see, for example, Ref.~\cite{Liu_2012} for an in-depth dispersion analysis of locally resonant beams).~For free wave motion, i.e., waves characterized by complex frequencies and either real or complex wavenumbers depending on the analysis method employed~\cite{Frazier_2016}, a corresponding wavenumber-dependent damping ratio is available, as shown in~Fig.~\ref{fig:Fig3}B.~The damping ratio is a direct measure of the strength of temporal attenuation of the propagating wave (see information on Bloch wave analysis in the Methods section).\\
\begin{figure}[b!]
	\begin{center}
		\includegraphics{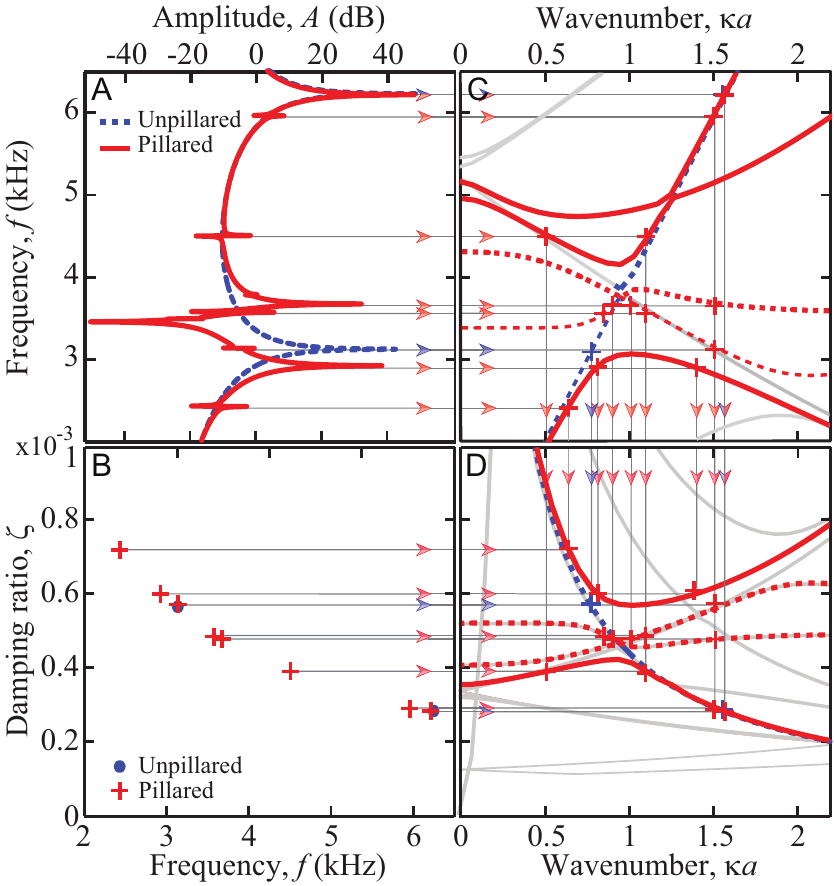}
		\caption{\small{\textbf{Calibration of unit-cell damping model via finite-structure modal analysis.}~(\it{A}\rm) Numerical FRF for the unpillared (dashed blue) and pillared (red) finite beams.~(\it{B}\rm) Modal-damping ratios for the finite beams obtained using a circle-fit technique.~Unit cell~(\it{C}\rm) dispersion diagram and~(\it{D}\rm) damping-ratio diagram.~For clarity, only the longitudinal modes are highlighted; the other modes are shaded. }}
		\label{fig:Fig4}
	\end{center}
\end{figure}

\indent \bf Experimentally-fitted Bloch damping model \rm Numerous studies on phononic materials have established a relation between dispersion analysis of the unit cell and eigenanalysis, the FRF, or transient response of a structure formed from a finite repetition of the unit cell~\cite{Hladky-Hennion_2005,hussein_dispersive_2006,Babaa_2017a,Babaa_2017b}. For example, localized modes are shown to appear, possibly inside band gaps, due to the truncation of the periodicity. However, no infinite-finite media correlation has been shown between the Bloch damping-ratios and the modal damping ratios for a corresponding finite system. We show here a novel procedure for establishing a prescribed damping model using experimental FRF data for a finite structure with the target of producing validated infinite-system damping ratios.~This procedure allows us to accurately predict dissipation levels within the dispersion analysis framework.~Using a standard experimental modal-analysis technique on the finite structure (see Figs.~\ref{fig:Fig4}A and~\ref{fig:Fig4}B and Supplementary Information), we select an initial set of damping parameters \(p\), \(q\) and \(\mu\) and solve for the complex eigenfrequencies.~Using these parameters, and a Bloch transformed version of the model given by~\eqref{EOM} as explained above, we obtain a damped dispersion diagram and a corresponding damping-ratio diagram~\cite{hussein_damped_2013,frazier_viscoelastic_2015}. We then iteratively establish a direct correlation between the vibration and the wave-propagation problems using, consecutively, the frequencies, wavenumbers, and damping ratios to link the models.~The iterative procedure is executed by implementing the following steps.~First, for each resonance frequency in the numerical FRF (Fig.~\ref{fig:Fig4}A), we obtain the modal damping ratios shown in Fig.~\ref{fig:Fig4}B using a circle-fit technique~\cite{ewins_modal_2000,kouroussis_easymod:_2012}. We then project each of these resonance frequencies onto the dispersion diagram (Fig.~\ref{fig:Fig4}C) and find the corresponding wavenumber(s).~Lastly, we project the modal damping values of Fig.~\ref{fig:Fig4}B onto the damping-ratio diagram (Fig.~\ref{fig:Fig4}D).~These steps are repeated, with the damping parameters \(p\), \(q\) and \(\mu\) being adjusted at each step, until the modal damping ratios and the corresponding values on the damping-ratio diagram match.~With this technique, we close the correlation loop between the finite and infinite systems and are able to develop an experimentally anchored prescribed damping model that accurately predicts dissipation within the dispersion analysis framework.\\
\begin{figure*}[h!]
	\begin{center}
		\includegraphics{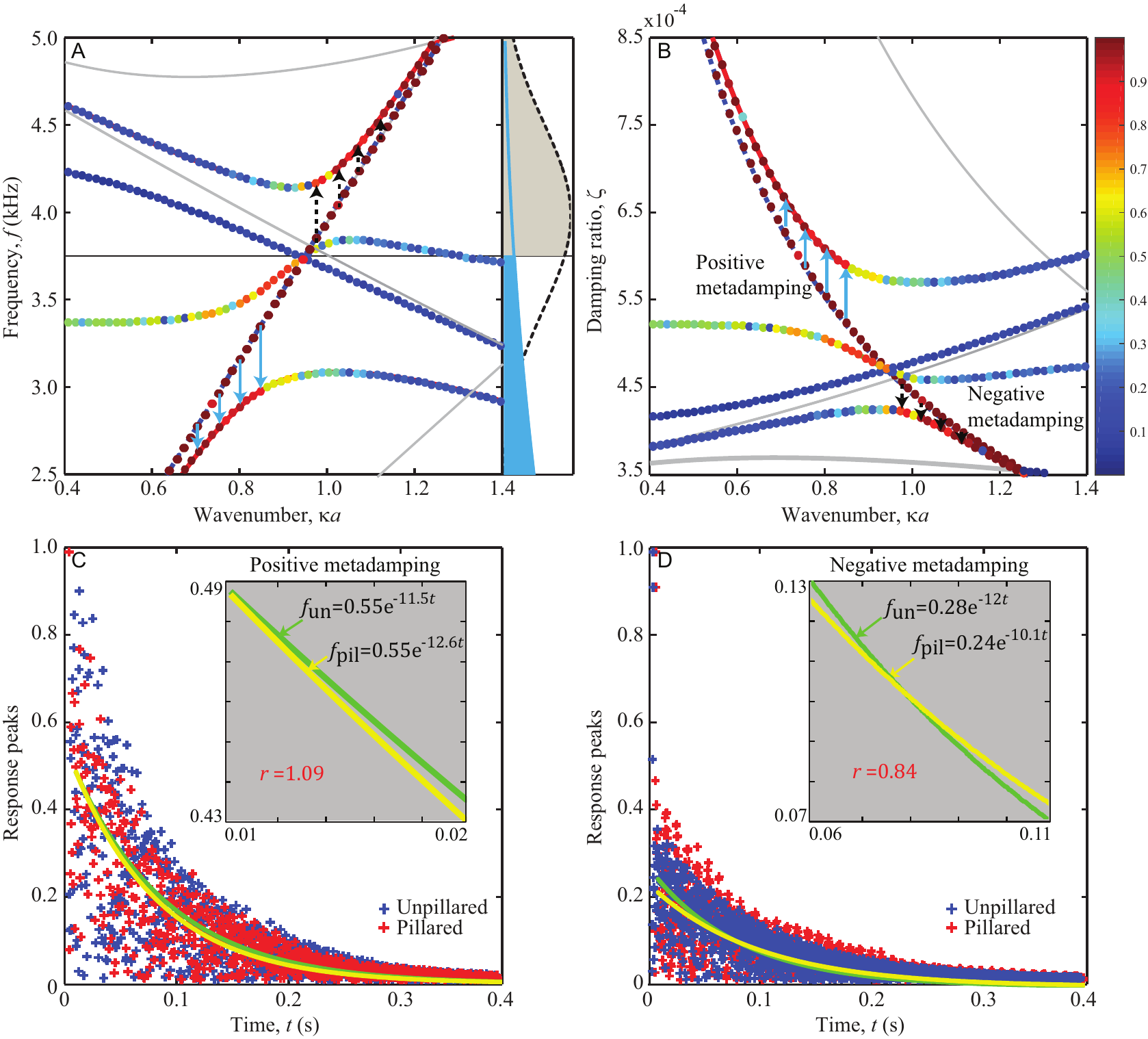}
		\caption{\small{\textbf{Theory of metadamping.}~(\it{A}\rm) Dispersion diagram with a close-up view of the hybridization region.~The color bar represents the degree of longitudinal polarization~\cite{achaoui_polarization_2010}: values range from 0 (pure shear) to 1 (pure longitudinal).~The inset shows the frequency content of the two impulse excitations.~The solid blue line [dashed black line] leads to positive [negative] metadamping.~(\it{B}\rm) Corresponding damping-ratio diagram.~(\it{C,D}\rm) Response peaks of the unpillared (blue) and pillared (red) time signals showing positive~(\it{C}\rm) and negative~(\it{D}\rm) metadamping.~The insets show a close-up of the fitted exponential curves (green for unpillared, yellow for pillared), and the metadamping ratio is given.}}
			\label{fig:Fig5}
	\end{center}
\end{figure*}

\indent \bf Theory of Metadamping \rm With our experimentally validated Bloch waves dissipation model in place, we now examine the concept of metadamping from a rigorous theoretical perspective.~Figure~\ref{fig:Fig5}A shows a zoom-in view of the dispersion curves plotted in Fig.~\ref{fig:Fig3}A, highlighting our region of interest that features resonance hybridizations.~We see that the straight line with positive slope, which is for the unpillared configuration, hybridizes into two curved lines, which are for the pillared configuration.~The corresponding damping-ratio curves are plotted in Fig.~\ref{fig:Fig5}B, where the hybridization phenomenon is again clearly noticed.~The blue arrows in Fig.~\ref{fig:Fig5}B indicate an increase in the damping-ratio values, that is, an increase in dissipation due to the introduction of the pillars.~In contrast, the black arrows indicate a decrease in the dissipation.~The blue arrowed transition is representative of positive metadamping, and the black arrowed transition is representative of negative metadamping.~Both these transitions take place for longitudinal wave modes.~Mapping these damping-ratio curves back to the dispersion band structure in Fig.~\ref{fig:Fig5}A enables us to identify the specific frequency ranges associated with each regime.~The positive and negative metadamping regions are seen to coincide, roughly, with the frequency ranges 2.5-3.75 kHz and 3.75-5 kHz, respectively. \\

\indent To demonstrate the effects of these metadamping properties on the temporal response of a corresponding finite structure, we separately apply two impulse excitation signals, each with a frequency content concentrated mostly within the respective metadamping frequency window mentioned above.~The Fourier spectrum of each of these signals is shown on the right side of Fig.~\ref{fig:Fig5}A, with the portions overlapping with the 2.5-3.75 kHz and 3.75-5 kHz frequency ranges shaded in blue and gray, respectively.~The blue excitation signal, which is the same as the one applied to produce the results in Fig.~\ref{fig:Fig2}D, is designed to trigger a positive metadamping response.~The black excitation signal, on the other hand, is tailored to trigger negative metadamping.~The temporally decaying maxima of the responses resulting from each of these two signals are shown in Figs.~\ref{fig:Fig5}C and ~\ref{fig:Fig5}D, respectively.~The results in Fig.~\ref{fig:Fig5}C are identical to those of Fig.~\ref{fig:Fig2}D, but plotted differently.~The metadamping ratio for each of these cases is~\(r_{\rm}=1.09\) (positive) and~\(r_{\rm num}=0.89\) (negative), respectively.\\

\indent To determine how the spacing of the pillars affects metadamping (positive and negative), we perform a parametric study and vary the unit-cell length \(a\) while keeping the pillar size constant.~In this analysis, we define two metrics, \(D_p\) and \(D_n\), to quantify the degree of positive and negative metadamping, respectively.~These metrics are evaluated by performing calculations on the damping ratio diagrams for the unpillared and pillared cases (see details in Supplementary Information).~Figure~\ref{fig:Fig6} shows the evolution of the metadamping metric as a function of the unit-cell length \(a\), for the positive and negative metadamping cases.~The value of the metric is shown to generally follow a decreasing trend for increasing unit-cell lengths \(a\).~This is expected because for longer unit cells, the influence of the pillar local resonances on the underlying beam dispersion curves will be ``less dense'' than for shorter unit cells.~It is noteworthy that positive metadamping is affected by the pillar spacing more strongly, as \(D_p\) decreases at a higher rate than \(D_n\) with increasing \(a\).\\
\begin{figure}[t!]
	\begin{center}
		\includegraphics{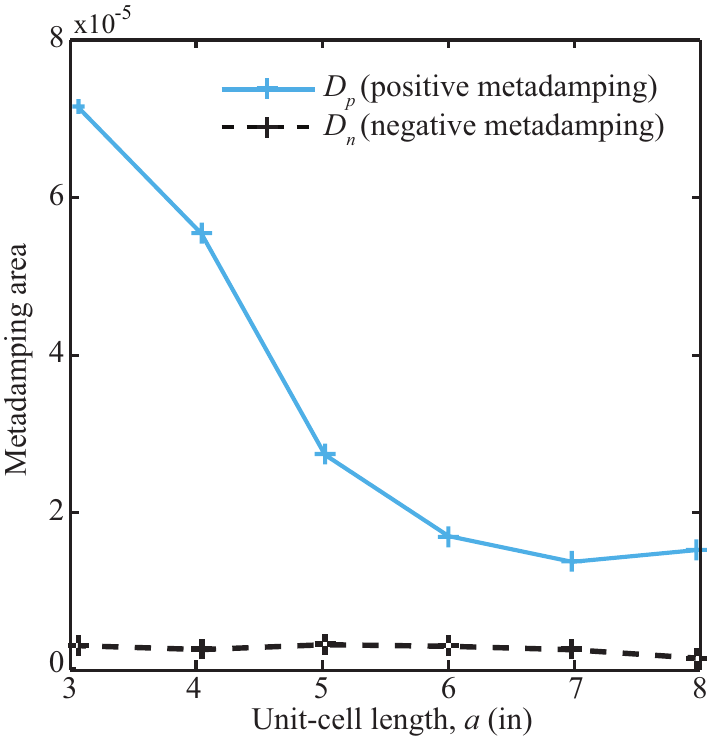}
		\caption{\small{\textbf{Parametric study:~Effect of pillar spacing on metadamping} Positive (blue) and negative (black) metadamping metrics as a function of the unit-cell length \(a\).}}
			\label{fig:Fig6}
	\end{center}
\end{figure}

\indent \bf Further insights by analytical model \rm To further characterize the metadamping phenomenon, we resort to an analytical model of a locally resonant rod admitting only longitudinal motion (which is the relevant mode of motion for all previous results).~In a recent study, Maznev~\cite{maznev_bifurcation_2018} used a simple model of an undamped locally resonant acoustic medium to analytically investigate a dispersion-resonance avoided crossing as a function of oscillator damping and oscillator interaction strength.~Here, we extend this analysis to determine the bounds on a prescribed damping parameter \(\eta\) for the locally resonant rod to exhibit positive or negative metadamping over the whole Brillouin zone (BZ).~A spring-mass oscillator is used to model the pillar.~The extension of our analysis from the framework of Ref.~\cite{maznev_bifurcation_2018} is straight forward as the equations of motion take the same form for both systems.~Because we are interested in metadamping, we compare the dissipation levels of the locally resonant rod with those of a reference uniform rod.~Unlike in Ref.~\cite{maznev_bifurcation_2018}, we assume that both the rod and the oscillators are damped.~Material damping is introduced in the rod through a complex Young's modulus \(E^*=E(1+i\eta)\), where \(E\) is the standard Young's modulus and is assumed to be frequency independent.~Deriving the dispersion relation for the damped rod yields the complex frequency \(\lambda_s\) in the form given by~\eqref{eigV} in the Methods section.~From \(\lambda_r\), we extract the wavenumber-dependent damping ratio \(\zeta_r(\kappa)=-\operatorname{Re}(\lambda_r)/\operatorname{Abs}(\lambda_s)\).~Note that because of our choice of the material damping model for the rod, \(\zeta_r\) is constant over the whole wavenumber range (unlike the case of the beam models analyzed earlier).~Similarly, we derive the dispersion relation for the locally resonant rod (the derivations are detailed in Supplementary Information) and obtain the two solutions \(\lambda_1\) (acoustic) and \(\lambda_2\) (optical).~From these two solutions, we extract the frequencies and the damping ratios \(\zeta_1\) and \(\zeta_2\).~The acoustic branch in this system is associated with positive metadamping; hence for positive metadamping to hold over the whole wavenumber range, the inequality \(\zeta_1(\kappa)-\zeta_r(\kappa)>0\) has to be satisfied for all \(\kappa\). Conversely, the optical branch in this model is associated with negative metadamping, thus the inequality for negative metadamping to hold over the entire BZ is \(\zeta_s(\kappa)-\zeta_2(\kappa)>0\).~We find that for an arbitrary choice of the coupling parameter \(\beta=0.01\), the condition for positive metadamping is true for any prescribed \(\eta\) and the condition for negative damping is satisfied for \(0<\eta \leq 0.14\) (see Supplementary Information for details).\\

\section*{Conclusions}
Metadamping is a dissipation emergence phenomenon whereby the level of dissipation may be enhanced or reduced in a metamaterial compared to a statically equivalent material with the same mass and type and quantity of prescribed damping.~Using a novel experimentally fitted damping model for accurate prediction of wavenumber-dependent dissipation, we provided evidence of metadamping in a finite~\it metastructure \rm setting with direct correlation to theory as described within the corresponding infinite~\it metamaterial \rm dispersion framework.~Both positive and negative metadamping have been demonstrated.~Our results show that either an increase or a decrease in dissipation, beyond nominal levels, is realizable by the proposed concept.~In both cases, metadamping may in principle be tailored to desired frequency ranges.~While elastic metamaterials are known to provide strong spatial attenuation inside band gaps, this work demonstrates the ability to exhibit strong, or weak, temporal attenuation.~Future work will extend the phenomenon of metadamping to the microscale and to waves driven at prescribed frequency.\\




\section*{Acknowledgments}
This research has been partially supported by the National Science Foundation CAREER Grant No. 1254931 and Grant No. 1538596.~The authors thank Professor Srikanth Phani for fruitful discussions.


\appendix

\renewcommand\thefigure{\thesection \arabic{figure}} 
\renewcommand\thetable{\thesection \arabic{table}}




\renewcommand{\thefigure}{A\arabic{figure}}
\setcounter{figure}{0}
\renewcommand{\theequation}{A\arabic{equation}}
\setcounter{equation}{0}

\section*{APPENDIX}

\section*{Beam samples}
The pillared beam is composed of four periodic aluminum unit cells each of dimensions \(8\times 1\times 1\) in along the $x$-, $y$-, and $z$-directions, respectively, with each unit cell featuring of a square pillar of dimensions \(0.5\times 0.5\times 2\) in placed on top of the base beam structure.~In order to solely focus on the effects of the pillar on dissipation and avoid effects stemming from the nature of the bonding, the pillars are not attached to the main beam by adhesion.~Instead, the pillared beam was manufactured by milling a thicker beam of dimensions \(32\times 1\times 3\) in using a CNC milling machine. \\

\section*{Finite-element numerical model}

In our finite-element (FE) model, we assume a density of \(\rho = 2700\) kg/m\(^3\), a Young's modulus of \(E = 68.9\) GPa, and a Poisson's ratio of \(\nu = 0.33 \).~The FE mass \(\mathbf{M}\) and stiffness \(\mathbf{K}\) matrices in Eq. (1) are obtained using three-dimensional 8-node brick elements (total number of elements in the unit cell are 4,096 for the unpillared beam and 4,352 for the pillared beam).~The damping matrix \(\mathbf{C}\) is constructed assuming proportional damping in the form \(\mathbf{C}=p \mathbf{M}+ q \mathbf{K}\).~The proportional damping parameters \(p\) and \(q\), as well as the relaxation parameter \(\mu\), are determined by the curve-fitting procedure described in the main article (see additional details in the following section).~Information on the FE implementation is available in Ref.~\cite{Hussein_PRSA_2009}; and information on the viscoelastic model adopted, and the application of Bloch's theorem to that model, are found in Refs.~\cite{hussein_damped_2013} and~\cite{frazier_viscoelastic_2015}.

\section*{Eigensolution}
\indent Following the treatment of general damping models by the state-space approach, which was originally developed for vibrations of finite structures (e.g.,~\cite{Newland_1989,wagner_symmetric_2003}) and recently extended to unit-cell wave propagation problems~\cite{hussein_band_2010,hussein_damped_2013,frazier_viscoelastic_2015}, we rewrite~\eqref{EOM} as a Bloch eigenvalue problem~\cite{bloch_uber_1929,Hussein_PRSA_2009}.~The solution of this problem gives us wavenumber-dependent complex eigenvalues~\(\lambda_s(\kappa)\) which physically represent the modes of damped-wave propagation:
\begin{equation}
\lambda_s(\kappa) =-\zeta_s(\kappa)\omega_s(\kappa) \pm i{\omega_d}_s(\kappa), \quad  s=1,...,n, 
\label{eigV}
\end{equation}
where $n$ is the total number of modes. The imaginary part \({\omega_d}_s\) represents the damped frequency corresponding to the \(s\)th Bloch mode, and the real part is the product of the wavenumber-dependent damping ratio \(\zeta_s\) and, in the case of Rayleigh (proportional) damping, the undamped frequency \(\omega_s\).~From~\eqref{eigV}, we obtain the damped dispersion diagram and the damping-ratio diagram.~The latter diagram provides us with the level of dissipation that each Bloch mode exhibits.\\

\section*{Experimental determination of damping parameters}

\subsection*{Description of the experimental set-up}
Experimental frequency response functions (FRF) are obtained as follows.~The beam under investigation is suspended using nylon cords in order to simulate free-free boundary conditions.~An accelerometer (PCB WJ35C65) is fixed on one of the ends of the beam, at the center point of the cross-section. Impulsive excitations are applied with an impact hammer (PCB 086C02) at a point located at the center of the opposite cross-section, such that only the longitudinal modes are excited and measured.
The measurements are collected with a NI-DAQ 9234 data acquisition system (the frequency rate is set to 25.6 kHz and the sample time is 5 s). The inertance spectrum is obtained by post-processing the time data using a commercial software package (MATLAB, The MathWorks Inc., Natick, MA, 01760, USA).

\subsection*{Initial curve fitting data}

As described in the main article, the experimental curve fitting of our numerical model is done in two steps.~In the initial step, a direct fitting is done to match the numerical FRF to an experimentally generated FRF.~In the second step, the model damping parameters are fine-tuned by correlating the finite and infinite problems using the proposed procedure demonstrated in Fig. 4 in the main article.~This ensures accurate prediction of wavenumber-dependent damping ratios.

\begin{figure}[t!]
	\centering
	\includegraphics{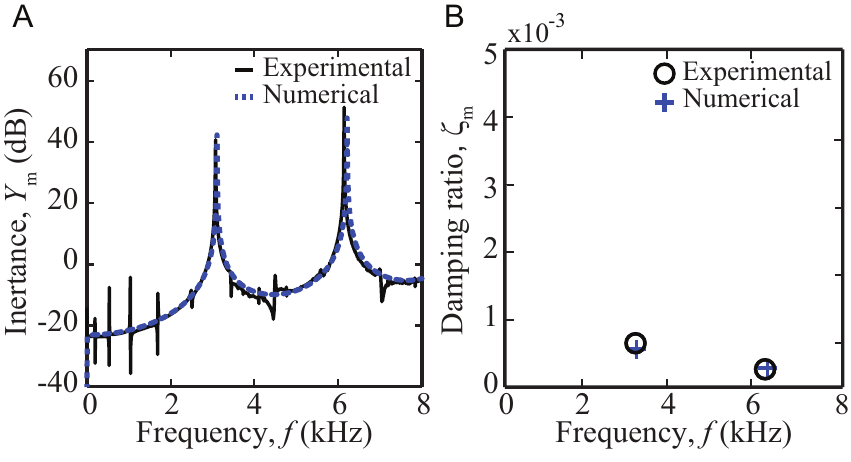}
	\caption{(A) Frequency response functions showing a comparison of experimental (solid black) and numerical (dashed blue) inertance spectra, and (B) experimental and numerical modal-damping ratio for the two shown resonances.~These results are for the uniform beam shown in Fig.~1}
	\label{fig:FigS1}
\end{figure}

Here we describe in the initial curve fitting step.~The modal damping ratios associated with the two measured resonance frequencies in the range of interest (0-8 kHz) are extracted using the circle-fit modal analysis technique.~The code EasyMod \cite{kouroussis_easymod:_2012}, which is available online, is employed for this purpose.~Using a trial-and-error approach, we vary \(p\), \(q\), \(\mu\) in the numerical model until the first two longitudinal modal-damping ratios match the experimental values.~Upon completing the second step described in the main article, prescribed damping values of \(p=22\), \(q=2.2\times10^{-7}\), and \(\mu=10^4\) (all in appropriate units) are found to produce a satisfactory match.~Figure \ref{fig:FigS1} compares the the numerical FRF and damping ratios with the corresponding experimental FRF and damping ratios using these damping parameters.
\\

\section*{Numerical simulation of finite structure and time decay characterization} 
The numerical time responses are obtained by implementing a direct time integration scheme for exponentially damped systems~\cite{adhikari_direct_2004}.~As in the laboratory testing, an excitation is applied at the center of the cross-section on one end of the beam along the longitudinal axis and the displacement on the opposite end of the beam is recorded.~The simulation is run for a total time of~\(t= 0.4\) s with a time step of~\(\Delta t= 3 \times 10^{-5}\) s.~For the positive metadamping case, an initial displacement in the form of a Gaussian impulse is applied:  
\begin{equation}
u_0(t)= e^{-\frac{(t-a)^2}{2b^2}}.
\end{equation} 
The constants are \(a= 0.01\) and \(b=8 \times 10^{-5}\); these values are selected to have the numerical excitation profile match the profile of the experimental impulse.~The frequency content of this profile falls mostly within the positive metadamping range of 2.5-3.75 kHz.~For the negative metadamping case, a prescribed force is applied that is also of Gaussian form:
\begin{equation}
f_0(t)= e^{-\frac{(t-a)^2}{2b^2}}e^{ict},
\end{equation} 
where \(a= 0.01\), \(b=3 \times 10^{-4}\) and \(c= 2.5 \times 10^4\).~Here, the constants are selected such that most of the frequency content falls within the negative metadamping range of 3.75-5 kHz.

\section*{Metadamping metric used in pillar-spacing parametric study} 
To study the effects of the pillar spacing on the degree of metadamping, we perform a parametric study and vary the unit-cell length \(a\).~Two metadamping metrics, \(D_p\) and \(D_n\), are defined to help quantify the amount of positive and negative metadamping, respectively, in the unit cell.~The procedure to obtain these metrics is as follows.~First, the Bloch modes are sorted using the Modal Assurance Criterion (MAC) criterion.\footnote{The MAC criterion is a measure of the degree of orthogonality between two vectors.~\cite{Allemang_2003}}~Then, the damping ratios corresponding to modes with dominant longitudinal motion are identified in the damping-ratio diagram.~This is done by computing the longitudinal polarization for each Bloch mode using a method described in Ref.~\cite{achaoui_polarization_2010}.~For the unpillared beam, we observe a single branch with dominant longitudinal motion; whereas for the pillared beam, two corresponding branches are extracted that exhibit longitudinal modes (the transition from one branch to two branches is a manifestation of the resonance hybridization phenomenon that takes place due to the presence of the pillars).~We will refer to these two branches as the \it upper branch \rm (associated with positive metadamping) and the \it lower branch \rm (associated with negative metadamping).\\
\begin{figure}[b!]
	\centering
	\includegraphics{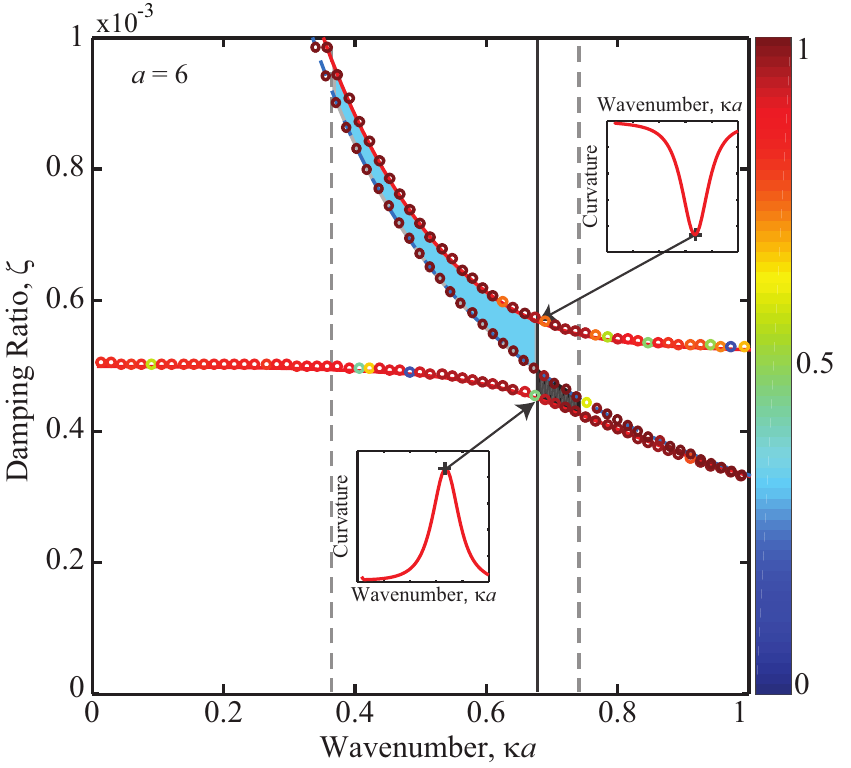}
	\caption{\textbf{Calculation of metadamping metric} Close-up on the relevant region in the damping ration diagram.~The left and right boundary lines (dashed grey) are drawn at the wavenumbers for which the difference between the unpillared and pillared damping ratios is 5\%.~The middle line is drawn such that it passes through the curvature maxima of the pillared branches (the curvature, or second derivative, for each branch is shown in the insets).~The value of the positive metadamping metric~\(D_p\) [negative metadampig metric \(D_n\)] is derived by calculating the area of the region shaded in blue [black].~The color bar indicates the level of longitudinal polarization; values range from 0 (pure shear) to 1 (pure longitudinal).}
	\label{fig:FigS2}
\end{figure}

As demonstrated in Fig.~\ref{fig:FigS2}, we define the positive [negative] metadamping metric as the area between the unpillared branch and the upper [lower] pillared branch, and between the left [right] and middle boundary lines.~The left and right boundary lines are drawn at the wavenumbers for which the ratio between the pillared and unpillared damping ratios differ by 5\%, i.e.,
\begin{equation}
\delta= \frac{|\zeta_{\rm pil}-\zeta_{\rm un}|}{\zeta_{\rm un}}\times 100 =5. 
\end{equation}
The middle line is drawn such that it passes through the extrema of the two pillared branches. The wavenumbers at which these extrema occur are determined based on the curvature (i.e., second derivative; see insets on Fig.~\ref{fig:FigS2}).~In order to ensure accuracy in the area calculation of the two metadamping regions, linear interpolation is performed along the wavenumber axis on the unpillared and pillared branches, between the left and right boundary lines. The areas are then calculated by slicing the two metadamping regions into quadrilaterals between consecutive wavenumbers and summing their areas. The formula used to find the area of each quadrilateral is as given by
\begin{equation}
\begin{aligned}
{\rm Area}= &\frac{1}{2} |(x_1y_2-x_2y_1)+(x_2y_3-x_3y_2) \\
& +(x_3y_4-x_4y_3) +(x_4y_1-x_1y_4)|
\end{aligned}
\end{equation}
where \(x_1=x_4=\kappa_i\) and \(x_2=x_3=\kappa_{i+1}\) are consecutive wavenumbers, \(y_1\) and \(y_2\) are the unpillared damping-ratio values, and \(y_3\) and \(y_4\) are the pillared damping-ratio values.\\

The demonstration provide in Fig.~\ref{fig:FigS2} is for a unit cell with length \(a=6\) in.~The positive metadamping region is highlighted in blue and the negative metadamping one in black.~The final results of the parametric study are summarized in Fig.~6 in the main article.
\begin{figure}[b!]
	\centering
	\includegraphics{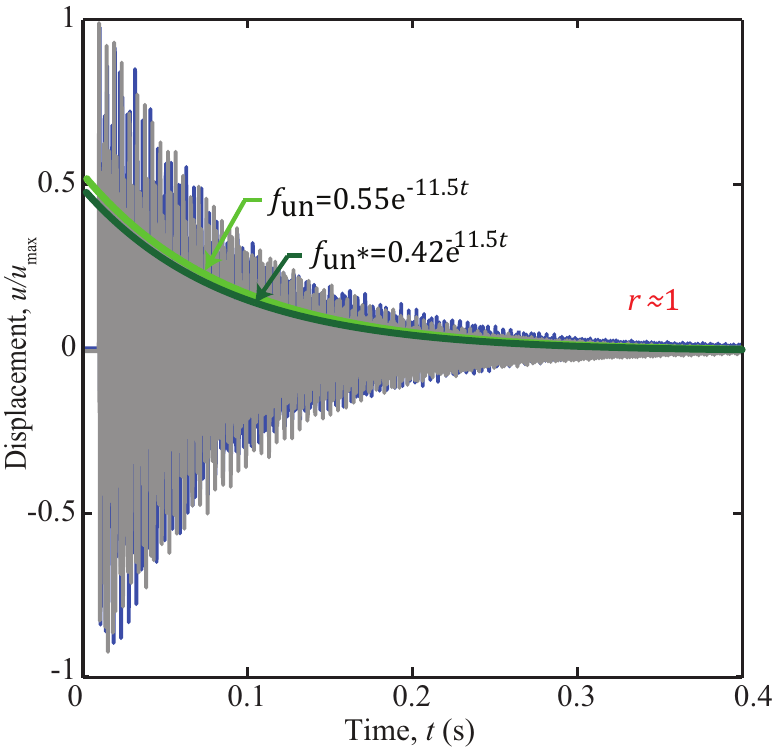}
	\caption{\textbf{Investigation of effect of added mass} Numerical time responses for two unpillared beams: the original uniform beam with volume \(V= 32\) in\(^3\) (blue) and another uniform beam with volume \(V= 34\) in\(^3\) (grey); the total volume of the latter unpillared beam is equal to the total volume of the pillared beam.~The light and dark green exponential curves are fitted to the time signals of the original unpillared beam (un) and the larger volume unpillared beam (un*), respectively.~The two signals are shown to exhibit the same time decay, thus confirming that the increase, or decrease, in dissipation is not due to change in volume of the damped material forming the structure.}
	\label{fig:FigS3}
\end{figure}
\section*{Effect of added mass on dissipation} 
We want to ensure that the added mass of the pillar is not a factor into our metadamping study.~The unpillared beam has a volume of \(V_{\rm un}= 32\) in\(^3\) whereas the pillared beam has a volume of \(V_{\rm pil}= 34\) in\(^3\).~Therefore, to show that the metadamping is due to the local resonance phenomena only and not due to the addition of extra mass that exhibits material damping, we perform a similar analysis as that of Fig.~2(D) in the main article on an unpillared beam that has the same volume as the pillared beam.~The numerical time responses and their curve-fitted exponential functions are shown on Fig.~\ref{fig:FigS3}.~The metadamping ratio, here redefined as \( r= b_{\rm un^*}/b_{\rm un}\), is shown to equal one; this confirms our hypothesis that metadamping emerges from the presence of local resonance and not from the addition of more damped material. 

\section*{Analytical characterization of metadamping}

\begin{figure}[b!]
	\centering
	\includegraphics{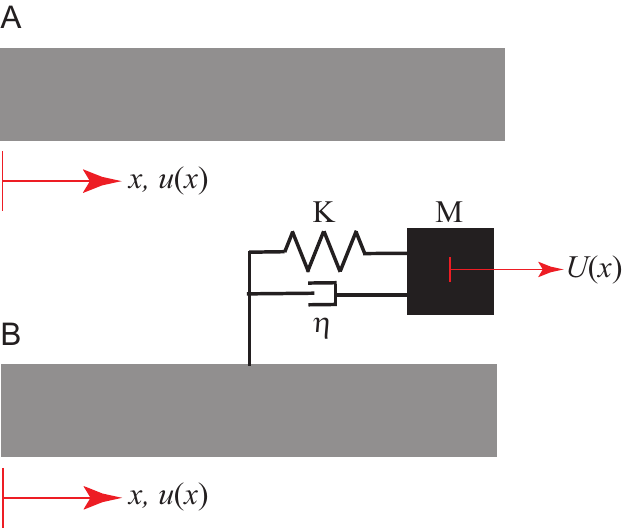}
	\caption{Schematic of (A) Simple rod, and (B) rod with a local resonator}
	\label{fig:FigS4}
\end{figure}

In this section, we follow the recent work by Maznev \cite{maznev_bifurcation_2018} to develop an analytical characterization of positive and negative metadamping.~To that end, we derive the dispersion relation for longitudinal motion in a thin homogeneous rod and in a rod with a spring-mass oscillator (see Fig.~\ref{fig:FigS4}), and compare their dissipation levels as a function of a prescribed damping parameter~\(\eta\).
The governing equation describing longitudinal displacements \(u\) in the rod is given by
\begin{equation}
\frac{\partial ^2 u}{\partial t^2}- (c_L^*)^2 \frac{\partial ^2 u}{\partial x^2}=0,
\label{EOM_rod}
\end{equation}
where \(c_L^*=\sqrt{E^*/\rho}=c_L\sqrt{(1+i\eta)}\) is the longitudinal speed of sound,~\(\rho\) is the density, and \(E^*=E(1+i\eta)\) is the complex Young's modulus.~For the locally resonant rod, an inertia term for the oscillator mass is added~\cite{maznev_bifurcation_2018} and \eqref{EOM_rod} becomes
\begin{equation}
\frac{\partial ^2 u}{\partial t^2 }+ M\frac{\partial ^2 U}{\partial t^2 }- (c_L^*)2 \frac{\partial ^2 u}{\partial x^2 }=0,
\label{EOM_LR_rod}
\end{equation}
where \(M\) is the mass of the oscillator and \(U\) is its displacement.~The volume fraction of the oscillator is negligible compared to the rod; therefore, the density and the modulus of the rod are not affected.~The equation of motion of the spring-mass oscillator is 
\begin{equation}
M\ddot{U}+ \eta(\dot{U}-\dot{u})+ K(U-u)=0,
\label{EOM_oscillator}
\end{equation}
where \(K\) is the spring constant and \(\eta\) is the damping coefficient. Note that the damping coefficient is chosen to be identical to the prescribed damping paramter (loss factor) for the supporting rod.~Assuming solutions of the form \(u=ue^{i\omega t-i\kappa x}\) and \(U=Ue^{i\omega t-i\kappa x}\), \eqref{EOM_rod}, \eqref{EOM_LR_rod} and \eqref{EOM_oscillator} respectively become
\begin{eqnarray}
(\omega^2-(c_L^*)^2 \kappa^2)u &=&0  \label{ODE1},\\
(\omega^2 \beta+\omega^2 -(c_L^*)^2+\kappa^2)u+ \beta \omega^2r &=&0 \label{ODE2},\\
\omega^2u+(\omega^2-i\gamma\omega -\omega_0^2)r &=&0,
\label{ODE3}
\end{eqnarray}
where \(r=U-u\) is defined as the amplitude of the oscillator displacement relative to the rod, \(\omega_0^2=K/M\) is the resonance frequency of the oscillator, \(\gamma=\eta/M\) is a term describing the damping in the oscillator, and \(\beta=M/\rho\) describe the strength of coupling.
From \eqref{ODE1}, we arrive at the dispersion relation for a damped rod:
\begin{equation}
(c_L^*)^2\kappa^2= \omega^2.
\label{dispersion_rod}
\end{equation}
Combining \eqref{ODE1} and \eqref{ODE3}, we obtain the dispersion for the damped rod with an oscillator:
\begin{equation}
(c_L^*)^2 \kappa^2=\omega^2(1 + \beta \frac{i\gamma \omega +\omega_0^2}{\omega_0^2- \omega^2+i\gamma \omega}).
\label{dispersion_LR}
\end{equation}	
Following the assumptions made in \cite{maznev_bifurcation_2018} that the damping is small, \(\gamma <<1\), and that the resonator has a small effect on the propagating waves, \(\beta<<1\), the term \(i\beta \omega\) in \eqref{dispersion_LR} can be discarded from the numerator:
\begin{equation}
(c_L^*)^2 \kappa^2=\omega^2(1 + \beta \frac{\omega_0^2}{\omega_0^2- \omega^2+i\gamma \omega}).
\label{dispersion_LR2}
\end{equation}
We set \(\omega_0\) and \(c_L\) to unity in \eqref{dispersion_rod} and \eqref{dispersion_LR2}, which respectively become:

\begin{eqnarray}
(1+i\eta)\kappa^2 &=& \omega^2, \\
(1+i\eta)\kappa^2 &=& \omega^2 +\frac{\omega^2 \beta}{1-\omega^2+i\gamma \omega} \label{dispersion_LR3}.
\end{eqnarray}
Rearranging \eqref{dispersion_LR3} yields:
\begin{equation}
((1+i\eta)\kappa^2- \omega^2)(1-\omega^2+i\gamma \omega)= \beta \omega^2.
\label{dispersion_LR4}
\end{equation}
The left-hand side of \eqref{dispersion_LR4} describes the interaction of the oscillator with the propagating waves in the rod, whereas the right-hand side describes the interaction of the two modes.
In the work presented in \cite{maznev_bifurcation_2018}, the author is interested in a so-called \textit{exceptional point}, which can be described as the bifurcation point separating the strong-coupling from the weak-coupling regimes. Therefore, if the two modes are uncoupled, one can assume that \(\kappa=\omega=1\) in the vicinity of that mode. \eqref{dispersion_LR4} can be further simplified as:
\begin{equation}
(2(\sqrt{1+i\eta})\kappa-\omega)(2-2\omega+i\gamma)=\beta.
\label{dispersion_LR5}
\end{equation}
The solutions of this quadratic equation are given by
\begin{equation}
\omega_{1,2}=\frac{1}{4}(2(\sqrt{1+i\eta})\kappa+2+i\gamma \mp \sqrt{(2-2(\sqrt{1+i\eta})\kappa+i\gamma)^2+4\beta}).
\label{solution}
\end{equation}
According to \cite{maznev_bifurcation_2018}, the two modes coalesce at the exceptional point, requiring the square root term to be equal to zero. We rederive here the values taken by \(\gamma\) and \(\omega\) at that exceptional point for the case when the underlying medium also has damping:
\begin{equation}
\gamma=\frac{2\sqrt{\beta}}{1-2M},\ \kappa=1,\ \omega=1+\frac{i \gamma (1+2M)}{4}.
\end{equation} 

We now characterize the condition on~\(\eta\) for the locally resonant rod to exhibit positive
and negative metadamping across the wavenumber range covering the entire Brillouin zone (BZ).~Given our interest in damped free waves, we now assume a solution of the form \(u=e^{\lambda t-i\kappa x}\) for the rest of the analysis.~With this formulation, the solutions for the uniform and locally resonant rods are the complex frequencies \(\lambda_s=i\omega_r=i\kappa\sqrt{1+i\eta}\) and \(\lambda_{1,2}=i \omega_{1,2}\), respectively.~From Eq.~(2) in the main article, we extract the wavenumber-dependent damping ratio \(\zeta_s(\kappa)=-\operatorname{Re}(\lambda_s)/\operatorname{Abs}(\lambda_s)\). Note that because of our choice of material damping model, i.e. a frequency-independent hysteresis damping model, \(\zeta_s\) remains constant over the whole wavenumber range.~Specifically, we extract the damping ratio for the acoustic \(\zeta_1(\kappa)\) and for the optical \(\zeta_2(\kappa)\) branches.\\
\begin{figure}[t!]
	\centering
	\includegraphics{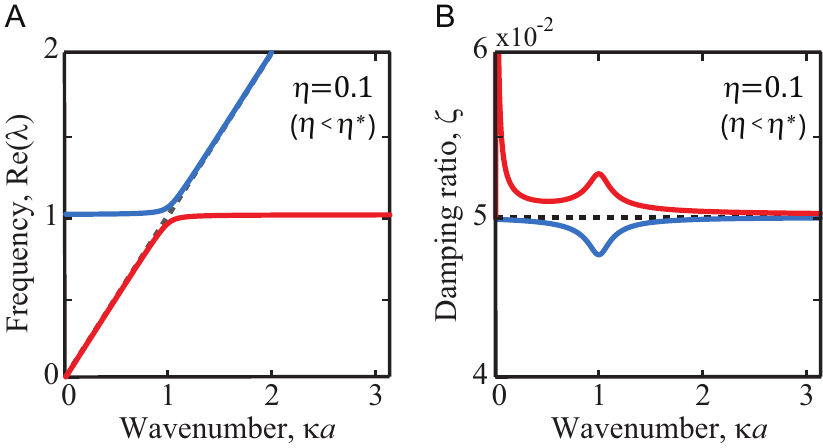}
	\caption{\textbf{Analytical dispersion analysis of locally resonant rod} (A) Damped dispersion curves for the uniform rod (dashed) and the locally resonant rod (solid).~The lower branch represents acoustic modes (red) and the upper branch represents optical modes (blue); (B) corresponding damping-ratio diagram.}
	\label{fig:FigS5}
\end{figure}
\begin{figure}[b!]
	\centering
	\includegraphics{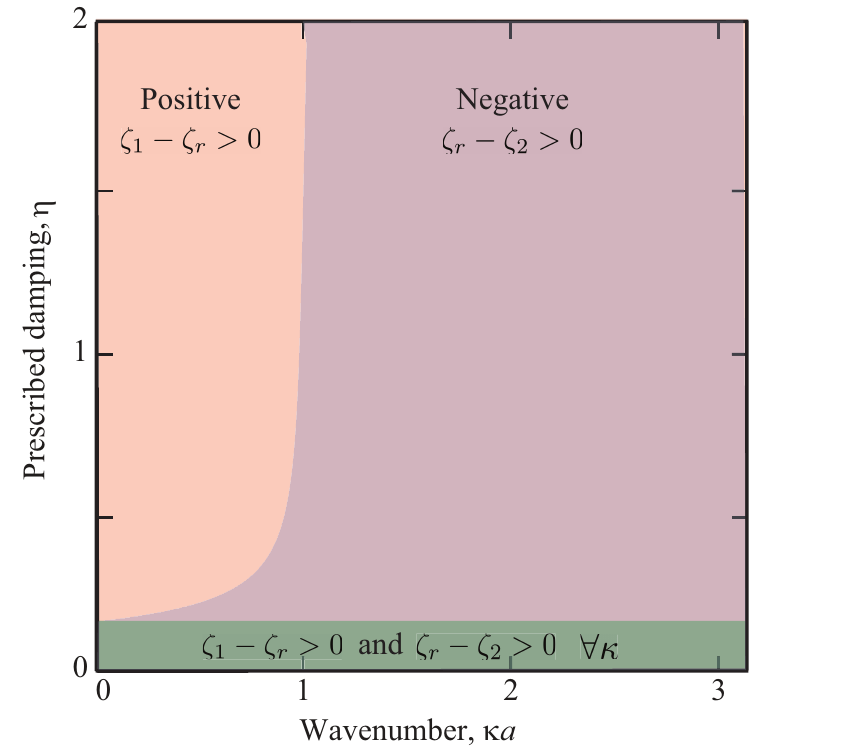}
	\caption{\textbf{Metadamping phase diagram showing bounds on prescribed damping} The region with inequality for positive metadamping is shaded in orange; and the region with inequality for negative metadamping is shaded in purple.~The region shaded in green represents the values of \(\eta\) that satisfy both positive and negative metadamping across the wavenumber range covering the entire Brillouin zone.}
	\label{fig:FigS6}
\end{figure}

We aim to determine the bounds on the prescribed damping value \(\eta\) such that the acoustic branch exhibits positive metadamping over the whole wavenumber, i.e., \(\zeta_1(\kappa)-\zeta_s(\kappa)>0\) is satisfied for \(0<\kappa<\pi\), and the optical branch simultaneously exhibits negative metadamping over the whole wavenumber range., i.e., \(\zeta_s(\kappa)-\zeta_2(\kappa)>0\) is satisfied for \(0<\kappa<\pi\). Using \eqref{solution}, we plot the damped frequencies \(\operatorname{Im}(\lambda_s)\) and \(\operatorname{Im}(\lambda_{1,2}) \) for prescribed values \(\beta=0.01\) and \(\eta=0.1\) (see Fig. \ref{fig:FigS5}A).~The hybridization caused by the local resonance occurs as expected around the resonance frequency of the oscillator \(\omega_0=1\). \\

In Fig. \ref{fig:FigS5}B, we show the corresponding damping ratios. The damping branch associated with the acoustic mode lies above the damping ratio branch for the homogeneous rod for the entire wavenumber range (positive metadamping). Conversely, the damping branch associated with the optical mode lies below the damping ratio branch for the regular rod for the entire wavenumber range (negative metadamping).~Due to the complexity of the closed form of both inequalities \(\zeta_1(\kappa)-\zeta_r(\kappa)>0\) and \(\zeta_r(\kappa)-\zeta_2(\kappa)>0\), we resort to numerical root finding to solve for the values of \(\eta\) meeting both conditions.~For an arbitrary value of \(\beta=0.01\), the results are summarized in the form of the phase diagram shown in Fig.~\ref{fig:FigS6}.~From this phase diagram, we find that in order to generate both positive and negative metadamping across the entire BZ, the bounds on \(\eta\) are such that \(0<\eta<0.14\).\\

\begin{figure}[t!]
	\centering
	\includegraphics{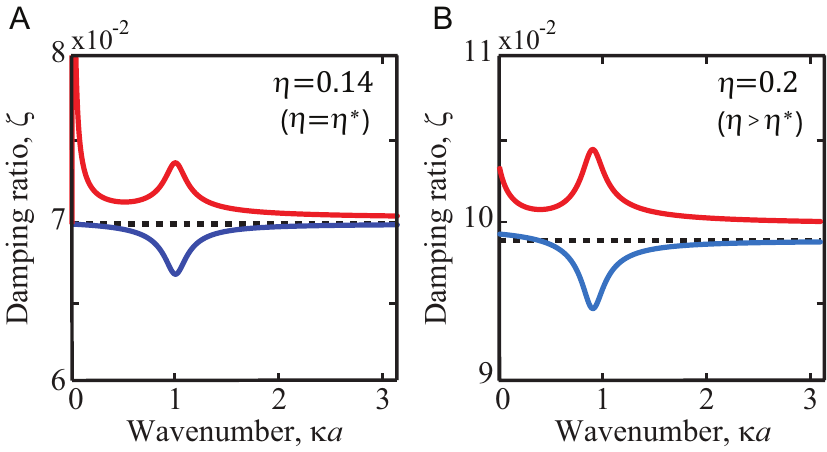}
	\caption{(A) Damping-ratio diagram for \(\eta=0.14=\eta^{*}\); (B) damping-ratio diagram for \(\eta=0.2>\eta^{*}\).}
	\label{fig:FigS7}
\end{figure}

Figure \ref{fig:FigS7} shows the damping ratio diagram obtained for two values of prescribed damping: \(\eta_1=0.1\) and \(\eta_2=0.2\).~In the first case (Fig.~\ref{fig:FigS7}A), \(\eta\) is less than the threshold value \(\eta_{max}\), which leads the optical damping branch to exhibit negative metadamping over the whole wavenumber range.~In the second case, however, \(\eta\) is greater than the threshold value \(\eta_{max}\).~The condition for negative metadamping is therefore no longer satisfied for \(0<\kappa<0.5\) where the optical damping branch is shown to lie above the homogeneous rod damping branch.\\

\bibliographystyle{ieeetr}
\bibliography{Biblio2}

\end{document}